\documentclass[11pt]{article}
\pdfoutput=1 
\usepackage{fullpage}
\usepackage{times}
\usepackage{multirow}
\usepackage{comment}
\usepackage{setspace}
\usepackage{graphicx}
\usepackage{natbib}
\usepackage{bbding}
\usepackage{pdflscape}
\usepackage{rotating}
\usepackage[font={small,it}]{caption}

\setcitestyle{authoryear,round,aysep={}}
\title{Teaching and Learning Data Visualization:\\ Ideas and Assignments}

\author{Deborah Nolan\\ Berkeley, CA 94720-3860\footnote{Deborah Nolan is Professor, Department of Statistics, University of California, 367 Evans Hall MC 3860, Berkeley CA 
94720-3860, (email: \texttt{deborah\_nolan@berkeley.edu})}\\
 Jamis Perrett\\Greely, CO 80639\footnote{Jamis Perrett is Product Analysis Lead, Monsanto, and
 Adjunct Associate Professor, University of Northern Colorado, Greely, CO 80639 (email: \texttt{jamis.j.perrett@monsanto.com}).}}

\begin{document}

\maketitle
\thispagestyle{empty}

\newpage
\setcounter{page}{1}

\begin{abstract}
This article discusses how to make statistical graphics a more prominent element of the 
undergraduate statistics  curricula.  
The focus is on several different types of assignments
that exemplify how to incorporate graphics into a course in a pedagogically meaningful way. 
These assignments include having students 
deconstruct and reconstruct plots, copy masterful graphs,  
create one-minute visual revelations, 
convert tables into `pictures', and develop interactive visualizations with, e.g., the virtual earth
as a plotting canvas. 
In addition to describing the goals and details of each assignment, 
we also discuss the broader topic of graphics and key concepts that we 
think warrant inclusion  in the statistics curricula.
We  advocate that more attention needs to be paid to this fundamental field of statistics at all levels,  
from introductory undergraduate through graduate level courses.
With the rapid rise of tools to visualize data, e.g., Google trends, GapMinder, ManyEyes, and
Tableau, and the increased use of graphics in the media, understanding the principles of good 
statistical graphics,
 and having the ability to create informative visualizations is an ever more important
aspect of statistics education.
\end{abstract}

\vspace*{.3in}


\onehalfspacing

\section{INTRODUCTION}

Over the past twelve years, we have increased the prominence of data visualization in our courses.
This began when we designed an upper division undergraduate course on computing for the 
statistics major that has a focus on data science.
In our experience, teaching graphics early in that course makes an easy entry point to
learning a statistics language such as R~\citep{bib:R}. 
Students are motivated to overcome the hurdles of learning computational thinking 
when the rewards are a creative expression of statistical findings.
We also advocate that a focus on graphics early in the course reinforces the importance of data 
visualization in data analysis, simulation studies, debugging, and all aspects of thinking with data.

We found that it is natural to include topics in visualization in the statistical 
computing course because of their computational nature. 
However, after developing this material and related assignments for the computing
course, we also found that a more rigorous treatment of visualization
can be taught at the introductory level, 
where traditionally only basic histograms, box plots, and scatter plots are introduced.
A statistical graph can offer an alternative
compelling approach to statistical thinking that focuses on important concepts
rather than procedural formulas.
We advocate that our quantitative reasoning courses  should address how to 
interpret and compose data visualizations, because these skills are parallel to the
communication skills taught  in core, introductory reading and composition courses 
offered at most institutions.

From these experiences, our perspective has evolved to 
thinking statistical graphics can and should play a larger role in all of our courses.
In this paper, our main contribution to this topic is to describe several assignments
that we have developed to: help students learn how to make
good graphs, incorporate the use of graphics in data discovery, and 
communicate effectively with graphs. 
We also think these assignments are effective in assessing a student's understanding
and competency in statistical graphics.
While many of these assignments were originally designed for the advanced 
computing course, they have been successfully adapted for the 
introductory course, and we believe they can be used successfully in a variety of other
statistics courses.

In Section~\ref{sec:principles}, we briefly describe the concepts and approach
we have taken for teaching graphics.  
Section~\ref{sec:assignments} provides examples of assignments that we have
developed and adapted for this purpose.
Briefly, these include assignments to 
deconstruct and reconstruct graphs, create one-minute visual revelations, 
copy masterful graphs, convert tables into `pictures', and design interactive visualizations.
In addition to assessing competency in data visualization,
these assignments also require students to demonstrate important skills related 
to the practice of statistics, such as communication and team work.

When we first introduced these visualization assignments in our courses,
we sought student feedback on them.
We developed evaluations aimed at ascertaining students' perceived benefits of
these homework assignments and projects.  
These evaluations are summarized in Section~\ref{sec:evaluation}.
As detailed there, students are generally very positive about the assignments.
Finally, in Section~\ref{sec:rubric}, we address the issue of evaluating student work.
We have developed an approach based on the literature for
assessing student competency. 
We have found that using this sort of rubric better aligns the grading of an assignment with its
purpose -- for students to demonstrate competency in core
statistical tasks such as how to analyze data and synthesize and present their
findings.

\section{STATISTICAL GRAPHICS IN THE CURRICULUM}\label{sec:principles}

Entire books are dedicated to the topic of statistical graphics, 
including several new and exciting contributions, e.g.,
\cite{bib:Cook}; \cite{bib:Murrell}; \cite{bib:Sarkar}; \cite{bib:Theus}; 
\cite{bib:Wickham}; and \cite{bib:Yau}. 
These monographs provide material for teaching data visualization at the advanced level.
Unfortunately, little attention has been paid to the principles of graphics 
in standard statistics texts. 
\cite{bib:gelmanRejoinder} asks: 
``If graphs are so great, why are they not more popular?'' 
and answers:
``Good statistical graphics are hard to do,
much harder than running regressions and making tables." 
We agree, and furthermore we think there are good reasons to 
teach data visualization more fully in both introductory and computing courses
and not delay until an advanced course where graphics is the sole topic of the course.

A common pedagogical approach in the introductory course is to learn about histograms
by drawing a few by hand. 
This approach offers a valuable means for understanding the concept of
a histogram, but it should not be the sole approach.
As \cite{bib:Cobb} explains, 
the field of statistics has developed under the constraints of available computational power.
In the past 10 to 20 years, computational capabilities have changed rapidly and so have the 
methodologies and practices of statistics. 
However, our teaching has not kept pace with these advancements.
(A notable exception is the recent effort to incorporate  resampling methods into 
introductory and advanced undergraduate courses, 
see e.g., \cite{bib:Chihara} and \cite{bib:Lock}.) 
Like many introductory offerings of statistics, we now employ software so students can
create histograms and other plots more readily.
We argue that we can and should also include computational advances in graphics
in these courses, 
such as smooth density curves and other local averaging and visualization techniques
for large amounts of data.

As we continued to teach and refine our data science/statistical computing courses, 
we synthesized the work of visualization experts,
including \cite{bib:Cleveland}, \cite{bib:Tufte}, 
\cite{bib:Tukey}, \cite{bib:Wainer}, and \cite{bib:WilkinsonBook}, 
into a two-week module on the principles of data visualization. 
In addition, we incorporated visualization into other topics of the course, thus
reinforcing the importance of statistical graphics as a tool for working with data.
We require students to use plots when debugging code and cleaning data.
In simulation studies, students summarize their findings in graphical form.
As another example, when students learn about XML-formatted data, 
we have them work with, e.g., KML to create visualizations
on Google Earth.
Furthermore, we advocate including a second programming language 
\citep{bib:NTL}  in order to provide a comparison 
language for students to abstract programming concepts from the syntax of the
specific language.  We have found JavaScript to work well in this regard, 
and it has the added bonus of enabling students to create dynamic Web visualizations.

Our goal in teaching statistical graphics is to provide a framework with which students can 
critique, compose and create graphs that usefully display information from data.
In all of our courses, we introduce a common vocabulary
for describing the elements of a statistical plot. 
We have used \cite{bib:Cleveland}
terminology with some success, which includes well known terms such as plotting symbol,
axis, axis label, legend, tick mark, and tick mark label, as well as, e.g., reference line and marker 
for adding auxiliary information to assist in understanding an important feature in a plot.

We have found that organizing the subject around the basic principles
of good graphics works well for courses at the introductory,
advanced undergraduate, and graduate levels. 
However, we do adjust the depth and focus on the topics according to the audience level. 
We present below the three guiding principles that we use in teaching this topic. 
These were distilled from the work of  \cite{bib:Cleveland}, 
\cite{bib:Tufte}, and \cite{bib:Wainer}.

\paragraph{1. Make the data stand out} The focus here is on revealing the structure of the data.
It includes discussion of how to fill the data region, transform data, 
choose an appropriate scale for an axis,  eliminate  chart junk and other superfluous material, 
and avoid having graph elements interfere with data, which includes topics such as 
over plotting, jittering, and transparency.  

\paragraph{2. Facilitate comparison} Here we focus on the questions: 
What is the important comparison? and How do we emphasize it?
Many of the same techniques for making the data stand out are considered from
this view point. 
In addition, we include notions of juxtaposing and superposing plots, and perception.
On perception, we address problems with: jiggling baselines, e.g.,
stacked bar plots;
representing one- two- and three-dimensions with length, area, and volume;
comparing lengths versus angles, e.g., a line plot vs. a pie chart; and 
selecting colors.  
With color, we touch on issues related to color blindness, luminance and saturation, 
and palettes for representing qualitative, diverging, and sequential values (see, e.g.,
\texttt{http://colorbrewer2.org}).

\paragraph{3. Add information} In addition to the usual conveyance of the importance of
labeling axes and using legends, we also discuss how to: use 
color and plotting symbols to convey additional information;
add context with reference markers and labels;
and write comprehensive captions that are self-contained, describe the
important features, and summarize the conclusions drawn from the graph.

\medskip 

In addition to these principles, we attempt to instill students with a few good work habits
related to creating statistical graphics and carrying out a data analysis, 
and we aim to convey some of the exciting challenges of modern visualizations. 
Examples of our approach to these additional issues follows.

\noindent
\textit{EDA and data types.} We typically introduce the topic of graphics 
through exploratory data analysis. In addition to the histogram, box plot, and 
scatter plot, we also describe less common but useful plots such as the rug plot, 
density curve, dot chart, bar plot, line plot, and mosaic plot. 
At the introductory level, the discussion of data types can help students choose an
appropriate type of plot for representing a variable or variables.  

\noindent
\textit{Integration throughout the data analysis cycle.}
After a thorough introduction to graphics via EDA, it is natural to include graphics
in other parts of the course.  
For example, in a data science course, 
we use graphics to check that we have correctly read in the raw data, e.g., 
we make visualizations to learn about the structure of the data, missing values, and outliers. 
We regularly check model assumptions through visualization, and
we have students summarize their findings with polished presentation graphics.   

\noindent
\textit{Iterative process.}
Creating a statistical graphic is an iterative process of discovery and fine tuning.  
We try to model this process in the
course by dedicating class time to an interactive iterative creation of a plot.
We begin either with a simple plot made by taking all the defaults or an ugly plot 
that screams for correction, 
and we transform it step-by-step, via a dialog with the students, into a graph that is data rich 
and presents a clear vision of the important features of the data. 

\noindent
\textit{Graphics for big data and designing unique visualizations.}
In addition to the histogram, box plot, and scatterplot, we also examine smooth versions
of these visual summaries, e.g., the density curve, violin plot, and smooth scatterplot, respectively.
The display of complex or large data, such as network topologies, DNA sequences,
and geographic maps, lend themselves to the development of specialized visualizations.
We provide examples of these and encourage students in advanced courses to 
develop their own approach to visualizing such complex data. 

\noindent
\textit{Interactivity.}
Today many Web-based technologies are available to display complex data in 
creative and imaginative ways.
Moreover, viewers of data visualizations have come to expect to be able to
interact with a graph on the Web by clicking on
it to get more information, to produce a different view, or control an animation.
We examine interactive visualizations in our courses through the framework of our
good-graphics guidelines, and in the more advanced courses,
we provide students with the tools to create them.

\section{EXAMPLES OF ASSIGNMENTS}\label{sec:assignments}

We have developed five graphics assignments
for teaching data visualization and computing. 
For each of these, we describe the assignment,
provide the motivation for the activity, and include an example visualization.
Table~\ref{tab:matrix5} summarizes the type of course
in which each assignment would be appropriate.
Most of the assignments are multi-purpose in that they can be adapted for 
many types of courses and levels of students.

\begin{table}
\small{
\begin{center}
\begin{tabular}{l | c | c | c |}
\multicolumn{1}{ c  }{ }  & \multicolumn {3}{c}{\textbf{Audience}} \\
 \multicolumn{1}{ c  }{\textbf{Assignment}} &  \multicolumn{1}{ c  }{Introductory} & 
 \multicolumn{1}{ c  }{Computing} & \multicolumn{1}{ c  }{Advanced Statistics} \\
\hline
\hline
Deconstruct - Reconstruct & \Checkmark & \Checkmark & \Checkmark \\
One Minute Revelation & \Checkmark & & \Checkmark \\
Table to Picture & \Checkmark & \Checkmark & \Checkmark  \\
Copy the Master & & \Checkmark & \\
Interactive Visualization &  & \Checkmark & \Checkmark\\
\end{tabular} 
\end{center}
\caption{This table provides a map of the types of courses in which each of the five types of 
graphics assignments can be used. The courses are organized into three types: 
introductory undergraduate statistics course;
data science course or a course in computational statistics; 
and an upper division undergraduate or introductory graduate statistics course.
Most assignments can be adapted for different types and levels of courses.
}\label{tab:matrix5}}
\end{table}

\subsection{Deconstruct and Reconstruct a Plot}\label{sec:deconRecon}
The process of identifying and deconstructing a problematic graph and then 
reconstructing it in a more appropriate form can help students better interpret,
critique, and construct meaningful graphics. 
With the deconstruct-reconstruct assignment, students have two major learning experiences.  
First, they have the opportunity to critique the appearance and application of a statistical graphic.  
Doing so can help their critical thinking skills.  
It can also help them see problems to avoid when constructing their own graphs.  
Second, students get the opportunity to create their own graph by modifying an existing graph
following the guiding principles (see Section~\ref{sec:principles}).

 \begin{figure}[ht]
 \centering
 \begin{tabular}{cc}
 \includegraphics[width=2.5in]{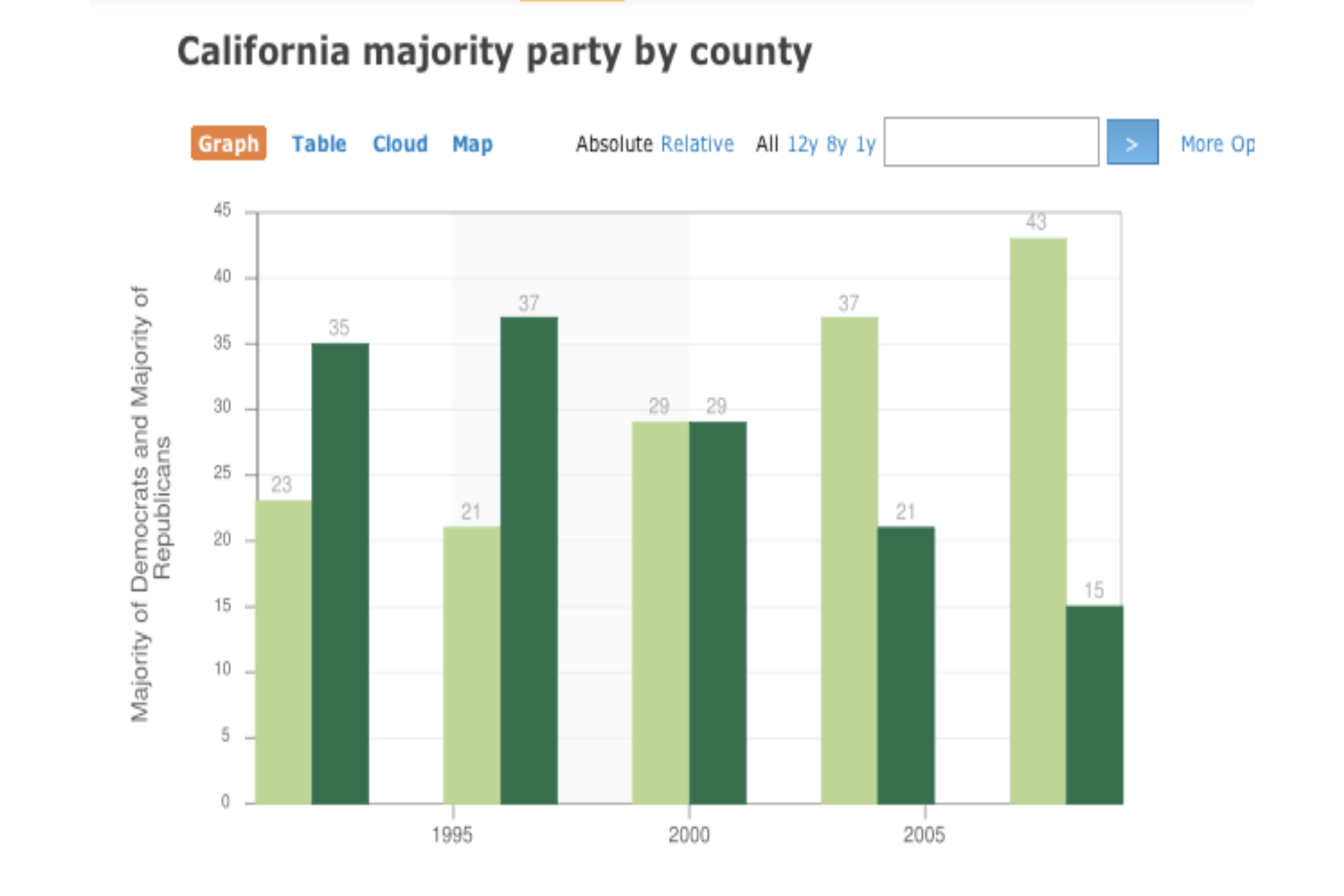} &
 \includegraphics[width=3.5in]{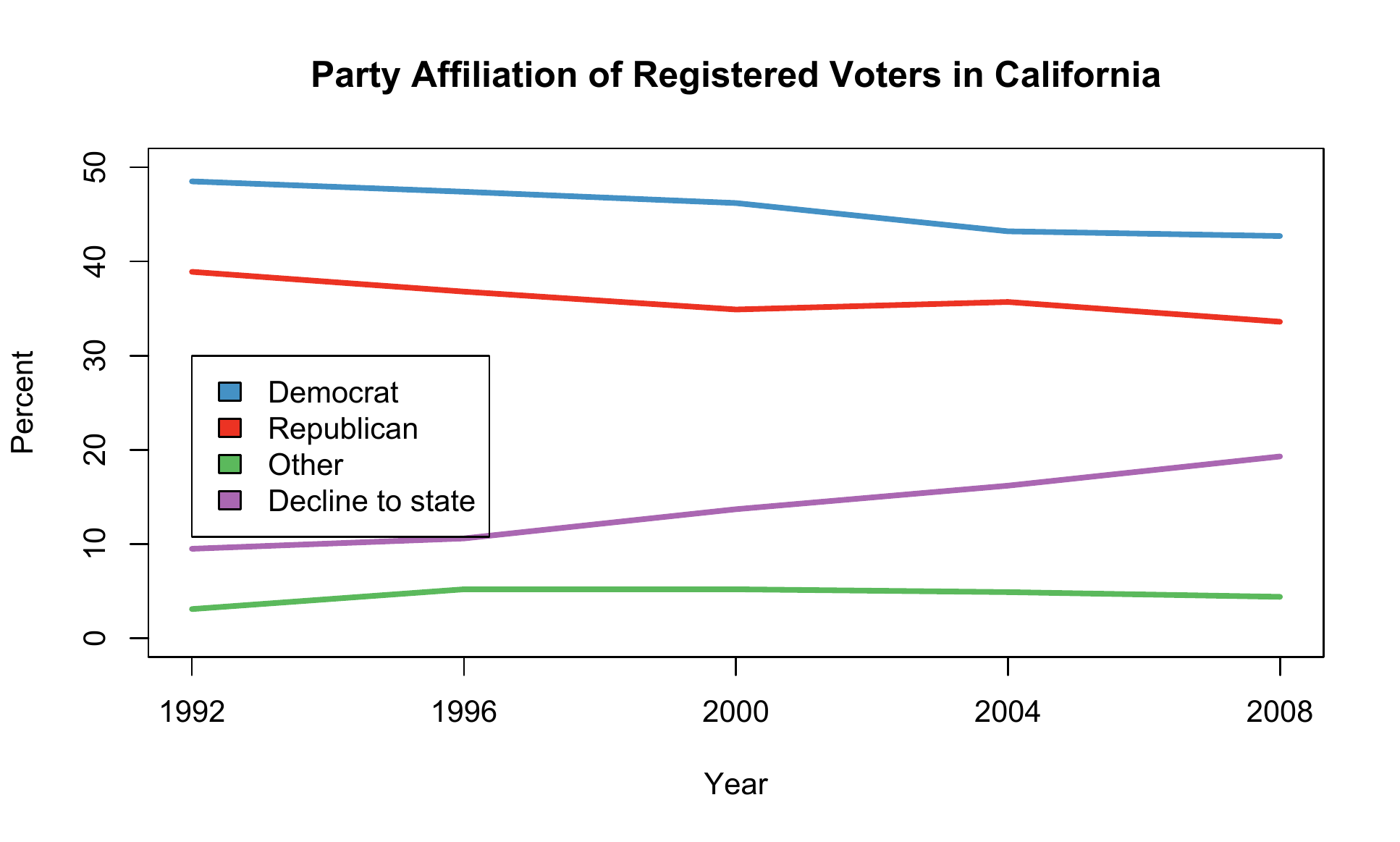}
 \end{tabular}
 \caption{The side-by-side bars in the plot on the left show the number of counties in 
California that have a majority of voters registered Republican (pale green) 
or Democrat (dark green).
Each pair of columns adds to 58, the number of counties in California. 
This plot was made on \texttt{swivel.com} (no longer a live site) 
from data available at the state of California's voter registration site.
The line plot on the left, compares the percentage of voters in California
registered with the Democrat (blue) and Republican (red) parties, as well as
those who decline to state an affiliation (black) or who are registered with 
another party (purple). Here, we see that the percentage of voters who decline to state an
affiliation has increased over the 16-year period, and the percentage of Republicans and 
Democrats have decreased roughly proportionally.}\label{fig:deconRecon}
 \end{figure}

We have found the ManyEyes Web site\\ (\texttt{www.ibm.com/software/analytics/manyeyes/})
to be a useful resource for this assignment. 
On Many Eyes, anyone can upload data and easily create a visualization.
There is a tremendous variety of data available on the site so students can find data on a topic
of interest to them.
Also, novices contributing to the site often make poor plot choices so there is plenty of 
opportunity for improvement.
Moreover, the data are available in CSV format for easy access.

After choosing a bad plot on a topic of interest to them, students     
must figure out the author's intended message 
and how it could be better conveyed in the plot. 
They identify all the basic elements of the existing plot using the 
graphics vocabulary,
and they critique the plot using our framework for evaluating plots. 
Their work includes writing a detailed caption for the ManyEyes plot that points out its 
deficiencies and describes what they think is the plot creator's intended message.

Students then remake the plot and fix the problems
that they have identified. This work can include,
e.g., adding more informative labels, transforming the data, etc.  
It can also include making an entirely different type of plot, 
if the original type is inappropriate for the data or if the message 
can be better conveyed with a different type of graph.
At this point, the students must also find additional data that enhances the message.
This additional information can be reference markers to aid in the interpretation of
important features in the graph or a new set of data to facilitate an important comparison.
Finally, they write a caption for the reconstructed plot that
points out the plot's main features and connects these features to the message 
to be gleaned from the graph.

The students work in pairs on this assignment, and we require each pair to
choose a different plot.
An example, appears in Figure~\ref{fig:deconRecon}.
The original plot (on the lefthand side) was completely overhauled into a line plot.
In the process of trying to remake the bar chart,
it was discovered that the data were incorrectly plotted. 
More importantly, it was decided that the message was not adequately
conveyed through county figures so new voter registration figures were acquired.
The resulting visualization (on the righthand side) tells a very different, and more 
accurate story of the change in voter registration.

We have been very impressed with the plots the students create.
After the assignment is due, we showcase some of the before and 
after plots that the students deconstructed and reconstructed. 
This serves as a second learning opportunity for how to construct informative plots, 
and it helps students calibrate their work.

We have used this assignment in introductory and advanced undergraduate courses.
At the introductory level, we often provide helper functions for transforming data. 
That is, we have found that the data often need to be transposed, 
variable types need to be specified, and new variables created from
variable names.  
We also allow the students to import these data into Excel and  ``fix'' them.
In the more  advanced courses, especially in courses with a
computational focus, we do not provide this assistance, and expect 
the students to create a reproducible workflow that includes transforming the
data into alternative formats conducive to making the new visualization.

\subsection{One-Minute Revelation}

Exploratory data analysis is an early step in the data analysis cycle that can uncover important
structure in the data and help guide the more formal analysis. 
This visualization assignment can help instill a work flow in
students' statistical practice by beginning with visual exploration of the data.
We typically use this assignment in an early stage of a group project.
Each member of the group must create a unique simple visualization of the data
and present a one-minute description of the graph and its relevance to the data analysis.
These plots and presentations are coordinated between team members so that
each plot offers a different issue.
Following each team's presentations, we have a discussion about how to use 
these revelations in determining next steps in the analysis.

These plots are expected to obey the principles of good graphics, but they are 
not meant to be formal presentation graphics. For example, they are not
expected to use additional information to make the plot information rich. 
The focus is on uncovering important features
in the data, synthesizing the various findings of the group into a coherent story,
and presenting a plan for further investigation.
 
This assignment helps students organize their workflow. 
They must start early on their project, 
The assignment reinforces the importance of effective team work, including  
communication between group members throughout all stages of the analysis.
It provides the opportunity for the students to present their ideas,
ask informed questions, and brainstorm with the professor about next steps.

We often include the grade for this assignment in the larger project grade.
And, we typically assign a score based on completion so that the students
are free to be creative in their exploration.

\subsection{Copy the Master}
A complex statistical graph can be considered a technological work of art, 
attracting the admiration of its viewers, remaining in their memory for years to come. 
We designed this assignment for use in computing-focused course,
and we originally had the following goals in mind. We wanted 
students to: a) gain practice with the statistical language;
b) discover the wealth of plotting functionality available in the language;
c) learn how to learn about new technologies, e.g., by taking advantage of online materials;
and d) create something `tangible' that could be built incrementally and viewed at each stage.
We were pleasantly surprised at the overwhelming positive response to the 
assignment and found that there were two additional goals that this assignment
achieved. Students also: 
e) learn to pay attention to the details in a visualization;
and f) show pride in creating a visualization as beautiful and complex as a master.  

We briefly describe three visualizations that we have used for this assignment.

\paragraph{Napoleon's March}
Minard's famous rendition of Napoleon's disastrous march in 1812-1813 
from France to Moscow and back incorporates many pieces of information
into a beautiful graphic\\ (see \texttt{http://en.wikipedia.org/wiki/Charles\_Joseph\_Minard}). 
The variables include the location and  size of the army.
The graphic also catalogs the date and temperature, and the direction of the army
is conveyed through color. Additionally, cities and rivers are used as reference markers along 
the path. 

\begin{figure}[ht]
 \centering
 \includegraphics[height=4in]{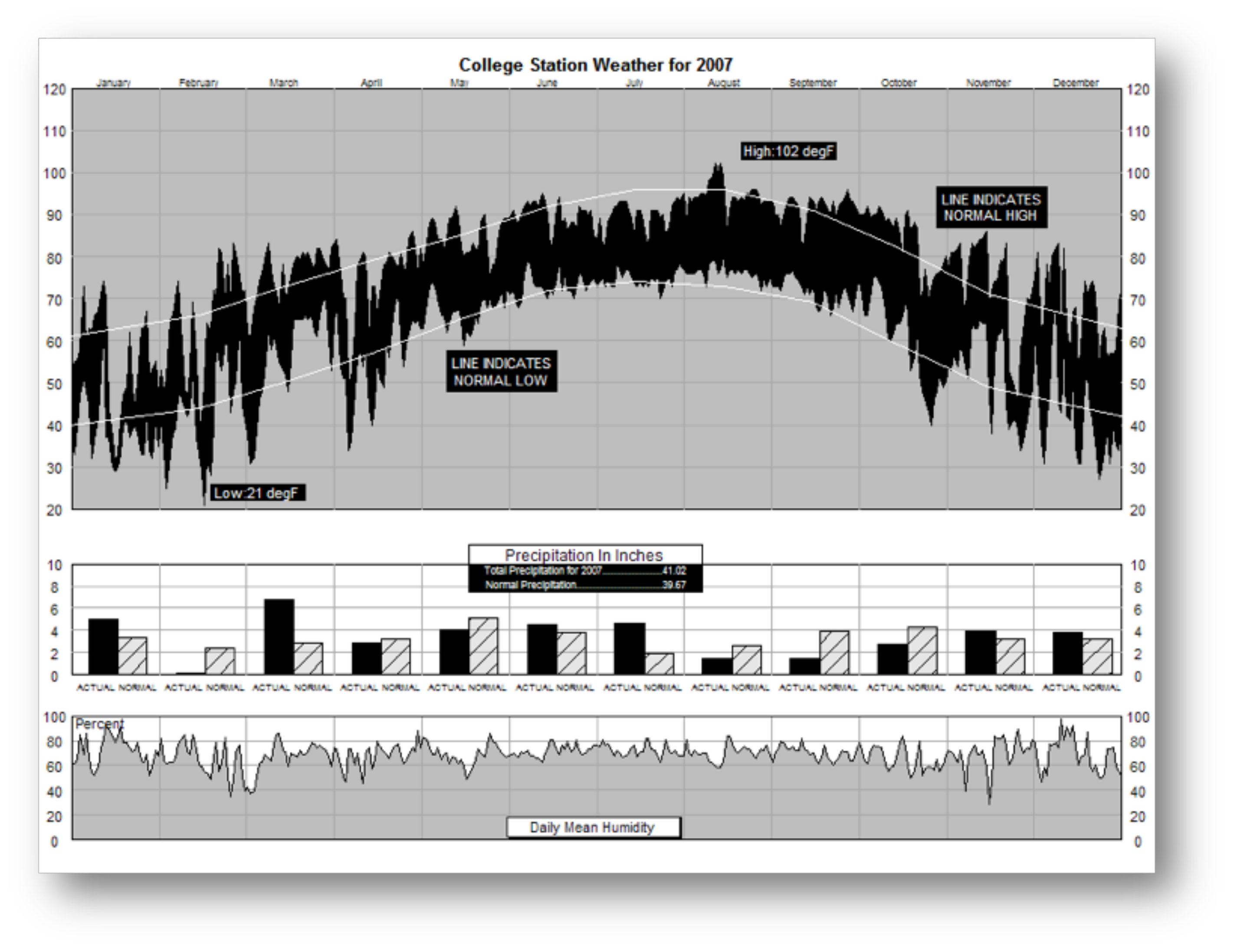}
\caption{This figure imitates the presentation of the 1990 NYC weather 
(NYT Jan. 11, 1981), but instead uses 2007 data from College Station, TX. 
 The graphic juxtaposes three plots, showing daily temperature (high and low) in the top plot, 
 monthly precipitation (actual and normal) in the middle plot, and daily percent humidity
 in the bottom plot.}\label{fig:CopyMaster}
 \end{figure}

\paragraph{New York 1980 Weather}
This graphic represents a comprehensive picture of the weather in New York City in 1980
(\textit{New York Times} Jan. 11, 1981).  
The variables include daily measurements of actual and historical normal  
temperature (high and low) and humidity, as well as monthly precipitation (actual and normal).
In this version of the assignment, 
we typically provide students with local current weather data. 
Now, students can't exactly duplicate the original plot. Instead, they are faced with
decisions about how best to display the data, 
e.g., the appropriate scales for the axes.
In some cases, we substitute a different weather variable and they must determine the most appropriate way to represent this information.
Figure~\ref{fig:CopyMaster} was created by a student in a graduate-level statistical 
programming course using the previous yearÕs weather data for College Station, TX.  
Using local weather data, provides added motivation,  
and gives students something to relate to in the assignment.   

\paragraph{Presidential Election Map}
The New York Times election map 
provides detailed county-level outcomes of the 2008 US presidential election\\
(\texttt{http://elections.nytimes.com/2008/results/president/map.html}).
Geographic patterns are revealed in this plot where the margin of victory
at the county-level is depicted with bubbles that vary in size according to the number 
of votes and are colored by the party of the winning candidate.
Transparency allows for bubbles in densely populated areas to overlap
without obscuring one another.
Similar to the weather example, we can provide students with results from other
elections, and have them create a plot that is comparable to the masterful New York Times graphic.

\subsection{Convert Tables into Pictures}
\begin{figure}[ht]
 \centering
 \includegraphics[height=3in]{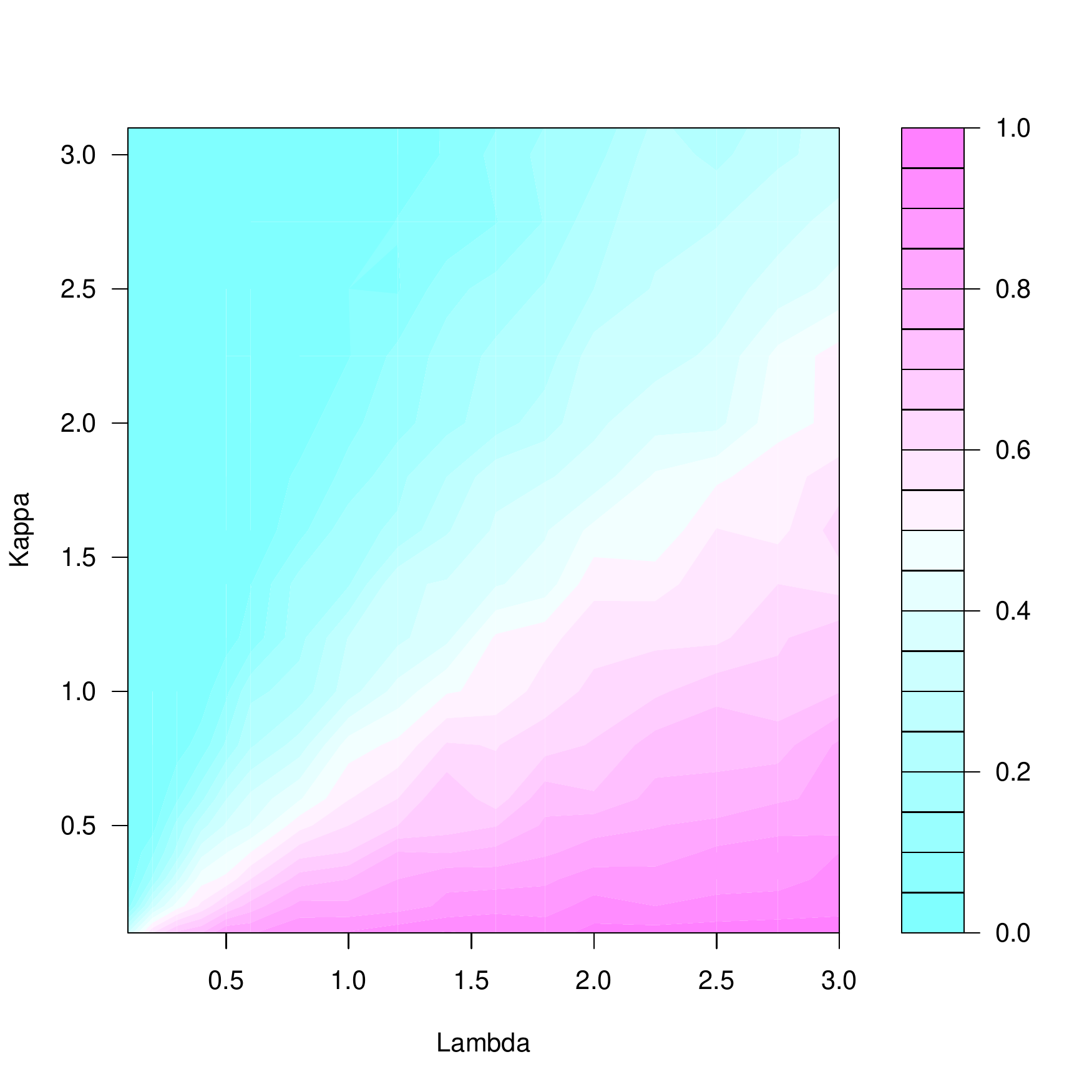}
 \caption{This image map presents the results of a simulation study of a birth-and-death process.
 Lambda and kappa are two parameters of the process, and the color denotes
 the proportion of simulations that survived at least 10 generations.
 The relationship between the two parameters is evident.}\label{fig:TableToPic}
 \end{figure}

In all of our courses, we use simulation studies to
understand the properties of estimators, stochastic processes, etc. 
For example, in the introductory course, we have students simulate the large sample 
behavior of an estimator other than the mean. 
In the computing course, we carry out more complex simulation studies of a 
stochastic process, such as a birth-and-death process. 
In all our simulation assignments,  students vary parameter(s)
values and simulate thousands of  times under each set of conditions.
Importantly, they are asked to summarize their findings graphically. 
Figure~\ref{fig:TableToPic} shows an image map of the results from a simulation study.  
The $x$ and $y$ axes provide the values of two parameters
of the process, and the color of the image represents a summary statistic of the
simulations at each $(x, y)$ pair. The relationship between the two parameters
is immediately apparent from this visualization. 
 
As noted by \cite{bib:GelmanPreach} ``it is still standard to display these latter 
summaries as tables rather than graphs.'' ... ``even though graphs are better suited 
for perceiving trends and making comparisons and predictions.'' 
We agree wholeheartedly and the goal of requiring a graphical summary
of the simulation results attempts to instill this practice.
In a related assignment, Horton (personal communication) has developed an assignment
for the introductory course. 
He asks students to find a table in Wikipedia on a topic of interest to them
and convert the table into a graph. 
Similar to the deconstruct-reconstruct assignment, the students need to think about 
the motivation for the Wikipedia table and how to display this information
graphically. 
This assignment has the additional feature of extracting data programmatically 
with Web scraping tools.

\subsection{Interactive Visualization}
\begin{figure}[ht]
\centering
\includegraphics[width=4.5in]{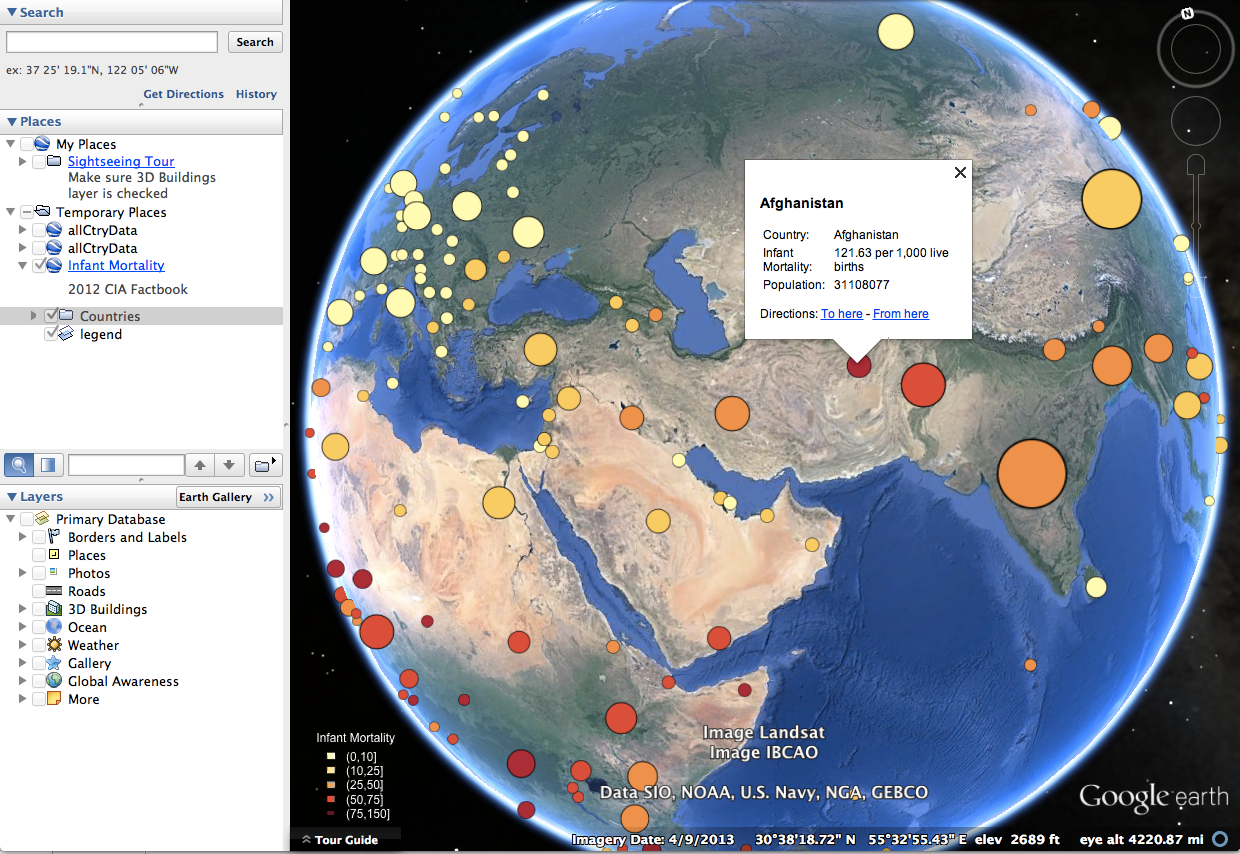}
\caption{This screenshot shows a visualization of infant mortality (color)
population (circle size) for countries in the CIA World Factbook.
The circles are located at the mean latitude and longitude of each country.
Additional information about each country is available in a pop-up window by clicking
on the country's circle.}\label{fig:IntViz}
\end{figure}

The role of graphics in the news has greatly expanded in the past few years.
For example, The New York Times graphics department includes statisticians as well as graphic artists,
and the graphics that appear in the online version of the news typically feature interactivity.
Indeed, some stories are entirely centered around a complex data visualization and
do not appear in print. 
Furthermore, applications such as Google Trends \citep{bib:GoogleTrends}, 
Google Earth \citep{bib:GoogleEarth} and GapMinder \citep{bib:GapMinder}, 
are changing the way business enterprises, researchers, and the general populous
visualize information.
Viewers expect to be able to interact with a plot on the Web in ways not 
previously possible.
Typically, regions in a plot contain hyperlinks to Web pages that contain 
more information on a subject, tooltips offer snippets of information about an
axis or data point, sliders control the animation of a graphic, and  
menus and buttons enable plots to be selected from a suite of visualizations.
The principles of good graphics can extend to this new domain, and 
for this reason, we include, the topic of creating interactive visualizations 
in our more advanced courses.
Figure~\ref{fig:IntViz} is an example of such an assignment.
Here, students create a visualization of data from the CIA World Factbook 
\citep{bib:CIAFactbook} for viewing in Google Earth. 
See \cite{bib:WilkinsonTech} for another example, 
where students analyze spatial/temporal data (the migration of an elephant seal) 
and present their findings via an animation on Google Earth.

\section{STUDENT FEEDBACK}\label{sec:evaluation}

We have carried out anonymous students evaluations of these assignments in several courses. 
At the introductory level, we asked students to provide general feedback on the use of R in 
the course and specific comments on the Deconstruct-Reconstruct assignment.
In advanced computing courses at the advanced undergraduate and graduate levels,
we asked students to comment on the usefulness and effectiveness 
of the Copy-the-Master assignment. 
This feedback has been used in formative assessment of the assignments.
We briefly summarize their feedback. 

\begin{figure}[ht]
\centering
\includegraphics[width=3in]{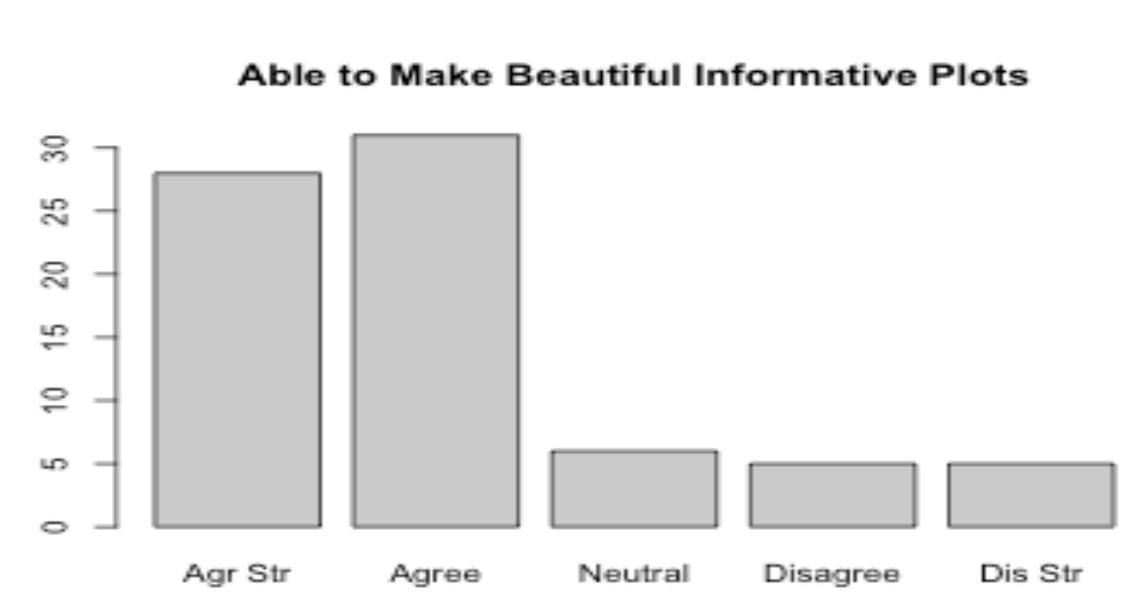}
\caption{This bar plot summarizes student response to one question from
an anonymous end-of-semester assessment of
the graphics portion of an introductory statistics class
(75 of 111 students responded).  58 (77\%) of the respondents 
agreed with the statement: ``with R they were 
able to make beautiful informative plots."}\label{fig:REval}
\end{figure}

At the introductory level, students were asked to evaluate the usefulness of the
computing aspect of the course. In particular, they were asked how strongly 
they agreed or disagreed with the following statement: 
with R, I was able to make beautiful and informative plots. 
Of the 111 students in the course, 75 responded to the evaluation. 
Their responses to this question are displayed in Figure~\ref{fig:REval}. 
The students overwhelmingly agreed or strongly agreed 
with this statement (58 of the 75, or 77\%) and 10 (13\%) disagreed or 
strongly disagreed.


\begin{figure}[ht]
\centering
\includegraphics[width=3in]{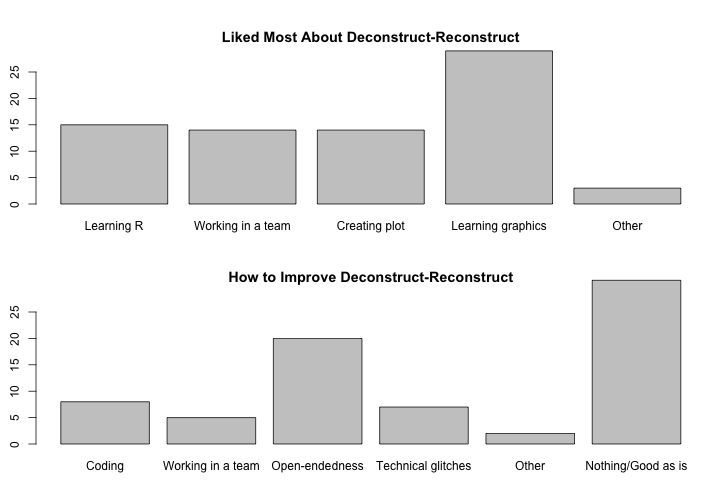}
\caption{In an anonymous end-of-semester assessment of
the graphics portion of an introductory statistics  class (with 75 of 111 students responding)
students were asked to provide reasons for what they liked most about
the Deconstruct-Reconstruct assignment and ideas for improving it.
Their responses about what they liked are well-aligned with the objectives of
the assignments (see Section~\ref{sec:deconRecon}). However, many students (20 of the 75)
wanted the assignment to be less open ended. The open-ended nature of the assignment
was also an objective of the assignment.}\label{fig:DeconReconEval}
\end{figure}

\paragraph{Deconstruct-Reconstruct} We used the Deconstruct-Reconstruct assignment
in this introductory course and asked students what they liked most about this assignment
and what suggestions they had for improving it.  Student responses are 
summarized in Figure~\ref{fig:DeconReconEval}.  The reasons provided to these
open-ended questions were categorized and tallied; if one student
provided more than one reason, then each was included separately in the tally.
As can be seen in the figure, the reasons provided for liking the assignment
are well-aligned with the assignment's learning objectives. Additionally, many
students stated they liked the assignment as is.  The biggest complaint about 
the assignment was that it was too open-ended. The learning objectives included
allowing students to exercise creativity in their plot making, which some students
found difficult (20 of the 75 respondents, or 27\%).

\paragraph{Copy the Master}
We asked for input about this assignment from students in an
upper division data science course with about 40 students
and from students in a Master's level statistical computing course of 25 students.
In the undergraduate course, the students copied Minard's rendition of Napoleon's march,
and in the Master's level course, they copied the 1990 NYT weather graphic.
In both courses, students' comments  describe this assignment's difficulty 
as a positive and rewarding aspect of the assignment.  
They also enjoyed the creativity of the assignment.  
For the undergraduate course, the following remarks were typical of the comments:\\
``I enjoyed the Napoleon's March [assignment] the most because it involved some creativity;"
 ``I felt [the assignment was] fun, and the results were very satisfying;"
 ``I had no idea that R can make such beautiful graphics;"
 ``I learned the most from project [Napoleon's March] because it was so complicated and time-consuming;"
 ``It felt like a task that was taken from start to finish, instead of inserting us in the middle like the homework. It was difficult, but rewarding in retrospect."

For the Master's students, the comments had a similar flavor:\\
``It was the most memorable assignment and challenging;"
``This is a more powerful way for me to learn programmingÉ  I liked the challenge;"
``I feel like I learned a great deal because of the fact that there was so much research involved;" 
``I learned so much and also felt much more confident about my ability using R after this assignment;"
``Very useful.  It challenged me to use all available resources (Internet, R Help, asking the professor for help, etc.) to get the program to work."
Furthermore, in response to closed questions about the assignment, 
18  of the 25 Master's students responded,
and all affirmed that the assignment helped them to learn basic R programming. 
Additionally, 16 of the 18 resounding agreed that researching a lot of the commands 
on their own was beneficial to the learning process,
and the same number of respondents appreciated the assignment involving real data.

\section{ASSESSMENT OF STUDENT WORK}\label{sec:rubric}
In evaluating student work we have tried several  different types of rubrics that we 
found in the literature or were suggested by colleagues. 
In our experience, rubrics based on assigning points to various aspects of the report, e.g.,
proper labeling axes in a plot, grammatically correct sentences,  
appropriate conclusions from an hypothesis test, etc., leads to a score that doesn't accurately
reflect whether or not the student has demonstrated mastery of the material, 
and this approach tends to promote quibbling over points. 
These problems occur even when we dedicate some of the points to a
holistic assessment of report.
On the other hand, we have found  that assignment of a single letter grade to
the student's work does not integrate well with our other forms of assessment, such as
exam scores.  

We have adapted a model for evaluating our graphics assignments and other data analysis
reports that takes a competency matrix approach.
The rows of this matrix correspond to different criteria that the student is expected to 
demonstrate in the assignment. 
The columns correspond to the extent of the student's competency in each of these criteria. 
The cells contain descriptions of how that level of competence is 
demonstrated. Table~\ref{tab:rubric} shows one example of a rubric we have used for 
evaluating a data analysis report.

\begin{sidewaystable}
\small
\begin{tabular}{p{4.5cm} | p{4.5cm}  p{4.5cm}  p{4.5cm} }
  & \multicolumn{3}{c}{ \textbf{Competency Level}}\\
 \textbf{Critical Task} &   \textbf{Needs Improvement} &  \textbf{Basic} &  \textbf{Surpassed} \\
 \hline
 \hline		
\textbf{Computation}
Perform computations &
Computations contain errors and extraneous code &
Computations are correct but contain extraneous/unnecessary computations	&
Computations are correct and properly identified and labeled \\
 \hline
\textbf{Analysis}
Choose and carry out analysis appropriate for data and context	 &
Choice of analysis is overly simplistic, irrelevant, or missing key component &
Analysis appropriate, but incomplete, or not important features and assumptions not made explicit &
 Analysis appropriate, complete, advanced, relevant, and informative \\
  \hline
\textbf{Synthesis}
Identify key features of the analysis, and interpret results (including context) &
 Conclusions are missing, incorrect, or not made based on results of analysis &
 Conclusions reasonable, but is partially correct or partially complete &
 Make relevant conclusions explicitly connected to analysis and to context \\
  \hline
\textbf{Visual presentation}
Communicate findings graphically clearly, precisely, and concisely  & 
 Inappropriate choice of plots; poorly labeled plots; plots missing &
 Plots convey information correctly but lack context for interpretation &
 Plots convey information correctly with adequate/appropriate reference information \\
  \hline
\textbf{Written}
Communicate findings clearly, precisely, and concisely &
 Explanation is illogical, incorrect, or incoherent. &
 Explanation is partially correct but incomplete or unconvincing	&
 Explanation is correct, complete, and convincing \\
 \hline
 \end{tabular}
 \caption{Example rubric for evaluating student work. A student is given a 
 \Checkmark in one of the three levels (needs improvement, basic,
 and surpassed) for each of the competencies (row in the table). This feedback is
 in addition to written comments.}\label{tab:rubric}
 \end{sidewaystable}
 
We have found that if the student is provided with detailed comments on his/her 
work and the completed matrix of competencies, then this evaluation creates a platform for
discussion between the instructor and student that is more process and
content driven than point driven.
If needed, we can convert the competency assessment into a numeric score. 
For example, we begin by giving a score of 85 for achieving basic competency in all 5 categories
shown in Table~\ref{tab:rubric}.
Then, we add points for competencies that surpass the basic level and subtract points for 
those that need improvement.
Typically, we add (subtract) 2 points for  competencies that 
have surpassed the basic (need improvement) for up to 3 competencies.
After that we add 4 points for a fourth competency that is surpassed
and 5 points for the fifth competency. 
In other words, it is increasingly challenging and rewarding to surpass the basic competency.
Similarly, we subtract 4 to 5 points for each competency beyond three that 
need improvement.
 
\section{SUMMARY}
This article is intended as a call to statistics educators to dedicate more course time to 
teaching statistical graphics.
A natural place for this topic is in the computing course \citep{bib:NTL} 
because computational skills are needed to create most of modern plots
and because it offers a rewarding venue for introducing a programming language.
However, we also advocate that more attention needs to 
be paid to graphics in introductory courses. 
Creating beautiful and informative statistical graphs can be very rewarding for students 
at this level as well as for the advanced student. 
Graphics also offers an alternative  venue for 
teaching core statistical topics and an opportunity to emphasize statistical thinking over calculations. 

We have presented several examples of assignments designed to
support integrating graphics into our curriculum.
These assignments are appropriate for courses that focus on statistics and on computing, and 
for audiences ranging from the introductory to graduate level.
Hopefully, the assignments have highlighted the key differences between an
assignment that simply expects the student to include a plot or two
as part of a data analysis and an assignment that focus on demonstrating competency in
creating and critiquing statistical graphs. 

\singlespacing
\bibliography{Graphics}

\begin{thebibliography}{23}
\providecommand{\natexlab}[1]{#1}
\providecommand{\url}[1]{\texttt{#1}}
\expandafter\ifx\csname urlstyle\endcsname\relax
  \providecommand{\doi}[1]{doi: #1}\else
  \providecommand{\doi}{doi: \begingroup \urlstyle{rm}\Url}\fi

\bibitem[{Central Intelligence Agency}(2015)]{bib:CIAFactbook}
{Central Intelligence Agency}.
\newblock The world factbook, 2015.
\newblock URL
  \url{https://www.cia.gov/library/publications/the-world-factbook/}.

\bibitem[Chihara and Hesterberg(2011)]{bib:Chihara}
L.~Chihara and T.~Hesterberg.
\newblock \emph{Mathematical Statistics with Resampling and R}.
\newblock John Wiley \& Sons, Hoboken, NJ, 2011.

\bibitem[Cleveland(1994)]{bib:Cleveland}
W.~Cleveland.
\newblock \emph{The Elements of Graphing Data}.
\newblock Hobart Press, Summit, NJ, 1994.

\bibitem[Cobb(2007)]{bib:Cobb}
G.~Cobb.
\newblock The introductory statistics course: A ptolemaic curriculum?
\newblock \emph{{Technology Innovations in Statistics Education}}, 1, 2007.

\bibitem[Cook et~al.(2007)Cook, Swayne, Buja, {Temple Lang}, and
  Hoffman]{bib:Cook}
D.~Cook, D.~Swayne, A.~Buja, D.~{Temple Lang}, and H.~Hoffman.
\newblock \emph{Interactive and Dynamic Graphics for Data Analysis: With R and
  GGobi}.
\newblock Springer, New York, 2007.

\bibitem[Gelman(2011)]{bib:gelmanRejoinder}
A.~Gelman.
\newblock Rejoinder.
\newblock \emph{Journal of Computational and Graphical Statistics},
  20:\penalty0 36--40, 2011.

\bibitem[Gelman et~al.(2002)Gelman, Pasarica, and Dodhia]{bib:GelmanPreach}
A.~Gelman, C.~Pasarica, and R.~Dodhia.
\newblock Let's practice what we preach: Turning tables into graphs.
\newblock \emph{The American Statistician}, 56\penalty0 (2):\penalty0 121--130,
  2002.

\bibitem[{Google}(2014{\natexlab{a}})]{bib:GoogleEarth}
{Google}.
\newblock Google earth software, version 6, 2014{\natexlab{a}}.
\newblock URL \url{http://www.google.com/earth/}.

\bibitem[{Google}(2014{\natexlab{b}})]{bib:GoogleTrends}
{Google}.
\newblock Google trends, 2014{\natexlab{b}}.
\newblock URL \url{http://www.google.com/trends/}.

\bibitem[Lock et~al.(2012)Lock, {Frazer Lock}, {Lock Morgan}, Lock, and
  Lock]{bib:Lock}
R.~Lock, P.~{Frazer Lock}, K.~{Lock Morgan}, E.~Lock, and D.~Lock.
\newblock \emph{Statistics: Unlocking the Power of Data}.
\newblock John Wiley \& Sons, Hoboken, NJ, 2012.

\bibitem[Murell(2006)]{bib:Murrell}
P.~Murell.
\newblock \emph{R Graphics. Computer Science and Data Analysis}.
\newblock Chapman Hall/CRC, New York, 2006.

\bibitem[Nolan and {Temple Lang}(2010)]{bib:NTL}
D.~Nolan and D.~{Temple Lang}.
\newblock Computing in the statistics curricula.
\newblock \emph{The American Statistician}, 64:\penalty0 97--107, 2010.

\bibitem[{R Development Core Team}(2012)]{bib:R}
{R Development Core Team}.
\newblock \emph{R: A Language and Environment for Statistical Computing}.
\newblock Vienna, Austria, 2012.
\newblock \texttt{http://www.r-project.org}.

\bibitem[Rosling(2008)]{bib:GapMinder}
H.~Rosling.
\newblock Gapminder: World, 2008.
\newblock URL \url{http://www.gapminder.org/world}.

\bibitem[Sarkar(2008)]{bib:Sarkar}
D.~Sarkar.
\newblock \emph{Lattice: Multivariate Data Visualization with R}.
\newblock Springer-Verlag, New York, 2008.

\bibitem[Theus and Urbanek(2009)]{bib:Theus}
M.~Theus and S.~Urbanek.
\newblock \emph{Interactive Graphics for Data Analysis: Principles and
  Examples. Computer Science and Data Analysis}.
\newblock Chapman \& Hall/CRC, New York, 2009.

\bibitem[Tufte(2001)]{bib:Tufte}
E.~Tufte.
\newblock \emph{The Visual Display of Quantitative Information}.
\newblock Graphics Press, Cheshire, CT, 2001.

\bibitem[Tukey(1977)]{bib:Tukey}
J.~Tukey.
\newblock \emph{Exploratory Data Analysis}.
\newblock Pearson, 1977.

\bibitem[Wainer(1997)]{bib:Wainer}
H.~Wainer.
\newblock \emph{Visual Relevations: Graphical tales of fate and deception from
  Napoleon Bonaparte to Ross Perot}.
\newblock Lawrence Erlbaum, Associates, Inc., Mahwah, NJ, 1997.

\bibitem[Wickham(2009)]{bib:Wickham}
H.~Wickham.
\newblock \emph{ggplot2: Elegant Graphics for Data Analysis. UseR}.
\newblock Springer, New York, NY, 2009.

\bibitem[Wilkinson(2005)]{bib:WilkinsonBook}
L.~Wilkinson.
\newblock \emph{The Grammar of Graphics}.
\newblock Springer, New York, NY, 2005.

\bibitem[Wilkinson(2010)]{bib:WilkinsonTech}
L.~Wilkinson.
\newblock The future of statistical computing.
\newblock \emph{Technometrics}, 50:\penalty0 418--435, 2010.

\bibitem[Yau(2011)]{bib:Yau}
N.~Yau.
\newblock \emph{Visualize This: The FlowingData guide to design, visualization,
  and statistics}.
\newblock John Wiley \& Sons, Hoboken, NJ, 2011.

\end{thebibliography}

\end{document}